\documentclass[structabstract]{aa}
\usepackage{txfonts}
\usepackage{natbib}
\usepackage{graphicx}
\bibpunct{(}{)}{;}{a}{}{,} 
\begin{document}

\title{3.3~$\mu$m PAH observations of the central kiloparsecs of  \mbox{Centaurus A}\thanks{Based on observations collected at the European Organisation for Astronomical Research in the Southern Hemisphere, Chile (Program 075.B-0618, PI: \mbox{Tacconi-Garman).}}}

\author{L.E. Tacconi-Garman\inst{1}
\and E. Sturm\inst{2}}

\institute{European Southern Observatory, Karl-Schwarzschild-Stra\ss e 2, 85748 Garching bei M\"unchen, Germany, \email{ltacconi@eso.org}
\and Max-Planck-Institut f\"ur extraterrestrische Physik, Giessenbachstra\ss e, 85748 Garching bei M\"unchen, Germany, \email{sturm@mpe.mpg.de}}
\date{Received date /
             Accepted date}
\abstract {} {The aim of this work is to further investigate the nature of PAH excitation and emission especially in the context of tracing star formation in a variety of extragalactic 
environments.  Here we turn our attention to the energetic environment of the closest AGN in our sample, Centaurus~A.\@} {Using ISAAC on the ESO VLT UT1 (Antu) we have made high spatial resolution 3.3\,$\mu$m imaging observations of the central kiloparsec of CenA.\@  These observations have been compared with star formation tracers in the near- and mid-infrared, as well as with mid-infrared tracers of nuclear activity.} {The nucleus is not devoid of PAH emission, implying that the PAH particles are not destroyed in the nucleus as might be expected for such a harsh environment.  However, we see the feature to continuum ratio decrease towards
the AGN\@.  As well, the 3.3$\,\mu$m PAH feature emission generally traces the sites of star formation in Cen\,A, but in detail there are spatial offsets, consistent with an earlier study of the starburst galaxies NGC\,253 and NGC\,1808. However, the feature-to-continuum ratio does not drop at the positions of star formation as was previously seen in that earlier study.  The cause for this difference remains uncertain.  Finally, our data reveal possible evidence for a nearly face-on, circular or spiral, dust structure surrounding the nucleus.
} {}
\titlerunning{PAH in CenA}
\authorrunning{\mbox{Tacconi-Garman} and Sturm}
\maketitle

\section{Introduction}
Over the last two decades it has become increasingly evident that emission from polycyclic aromatic hydrocarbons (PAHs) can be used as a powerful diagnostic of the physical conditions in a virtually unlimited number of different environments, from the environs of individual stars to globally-averaged entire galaxies \citep[e.g.][and references therein]{Peeters2004}.  There are prominent emission features at 3.3, 6.2, 7.7, 8.6, 11.3, and 12.7$\,\mu$m.  The strength of these features depend on the characteristics of the emitting particles \citep[size, geometry, composition, charge state, etc.; e.g.][]{Allamandola1989,Bauschlicher2008,Bauschlicher2009,Draine2001,Schutte1993}.  Thus, ratios of PAH emission at various of these wavelengths (especially to emission at 3.3$\,\mu$m) can be used as powerful tracers of grain size, grain photoionization state, etc.~\citep[e.g.][]{Bauschlicher2009}.  For the particular case of dusty galaxies, early studies with the {\em Infrared Space Observatory\/} \citep[e.g.][]{Genzel1998, Rigopoulou1999, Tran2001} used the presence and strengths of spectral features attributed to PAHs in comparison to ionic line and continuum emission to disentangle the relative contributions of starburst and active nuclei to the overall luminosity in samples of ultraluminous {\em IRAS\/} galaxies (ULIRGs).  Later, with the low resolution spectroscopy afforded by the {\em Spitzer Space Telescope\/}, it became not only possible to use the PAH emission features to estimate redshifts to IR bright galaxies to at least $z\sim2$ but also to perform more detailed studies of those galaxies, thereby placing constraints on the contribution of star formation and AGN to the cosmic IR background \citep{Schweitzer2006, Spoon2007, Teplitz2007, Valiante2007, Yan2007}.

A few key observational facts have emerged from these studies and many others covering a wide range of source types \citep[][to cite but a few]{Peeters2002, vanDiedenhoven2004, Galliano2008}.  First, PAH emission is seen in normal, starbursting, and active galaxies.  However, although there is little difference in the profiles of the PAH features from one galaxy to another, there are observed variations in the ratios of the PAH feature emission at different wavelengths \citep[e.g.][]{Peeters2004}.  Such ratio variations have been used as probes of the details of the PAH characteristics (e.g.~size and charge) in the emission regions  and/or of the local conditions \citep[e.g.][]{Bolatto2007, Bauschlicher2009, Haynes2010, Diamond-Stanic2010, Galliano2008, Haan2011}.

Secondly, there appears to be a threshold at a metallicity of 12$+\log(O/H)\sim 8.2$ below which PAH emission features appear much fainter than in higher metallicity systems \citep[][and references therein]{Engelbracht2005, Wiebe2011, Haynes2010}.

To investigate these issues in more detail we have undertaken an imaging study of the 3.3~$\mu$m PAH feature emission in a sample of nearby, IR bright, well-studied low-metallicity dwarf galaxies, starburst galaxies, and Seyferts spanning a range of physical environments.  High spatial resolution 3~$\mu$m imaging has been obtained for the entire sample, and the results for two starburst galaxies, NGC\,253 and NGC\,1808, have been published in \citet{Tacconi-Garman2005}. 

The results of that study show that although generally the PAH emission does peak in the inner star forming regions, the lack of detailed correlation (either positive or negative) between PAH emission and sites of recent star formation indicates the connection between the instantaneous star formation rate and PAH emission may not be so direct as advocated in the studies of nearby galaxies from large aperture work \citep[e.g.][and references therein]{Tran2001}.  This may signal that the PAH emission better traces the general B star population than sites of recent massive star formation \citep{Spoon2003,Boselli2004,Calzetti2010}.  Moreover, in those galaxies we found that the PAH feature-to-continuum emission actually decreases at the star formation sites \citep[see Fig.~6 and 7 in][]{Tacconi-Garman2005}, perhaps owing to mechanical energy deposited into the ISM, photoionization of the PAH, or photodissociation of the PAH\@.  In addition, in NGC\,253 we find the first evidence for PAH molecules in a starburst-driven galactic superwind.

In the present work we turn our attention to the more energetic environment presented by the nearest Seyfert galaxy, Centaurus\,A\@.  This galaxy is by far one of the most well-studied of all galaxies, and a comprehensive review of this work has been presented in \citet{Israel1998}.  The proximity of Cen\,A allows us a unique opportunity to separate the active galactic nucleus and jet from surrounding structures.  This, in turn, provides us with the ability to cleanly separate the impact that the nuclear activity has on the PAH feature emission.  Our observations and the techniques employed for data reduction are described in \S2 and a general description of our results is given in \S3.  In \S4 we put our results in context with those of other authors studying Cen\,A at other wavelengths, and in \S5 we provide our conclusions.

\section{Observations and Data Reduction}
\subsection{Observational Details}
Data on Cen\,A have been obtained in Service Mode with the infrared camera and spectrograph ISAAC on ESO's Very Large Telescope UT1 (ANTU) on 11/12 July 2005. ISAAC was operated in the LWI3 mode, with a pixel scale of 0.1478 arcsec/pixel, and a corresponding field of view of 151$^{\prime\prime}$ $\times $ 151$^{\prime\prime}$. We used the two narrow band filters NB\_3.28 and NB\_3.21, centered on the 3.28~$\mu$m PAH feature and the underlying continuum at 3.21~$\mu$m, respectively. The spectrum near the 3.3~$\mu$m PAH feature shows continuum on the blue side and a weaker PAH feature at 3.4~$\mu$m \citep{Sturm2000,vanDiedenhoven2004} making the use of a single blue side filter for continuum more accurate than if the feature were bracketed with two filters.  

In both filters we took a series of exposures, each consisting of a stack of 160 0.5\,second sub-exposures,  with the nucleus placed in turn near the middle of each of the four detector quadrants with jitter applied.  This strategy was adopted under the assumption that the region over which PAH and/or continuum emission would be detectable would be small enough not to fall in any of the other three detector quadrants which for the exposure in question did not contain the nucleus, making data in those quadrants viable for sky measurement purposes.  After frame selection the total exposure time was 53\,min and 57\,min for the NB\_3.21 and NB\_3.28 filters, respectively. The mean seeing at 3.3~$\mu$m was about 0.6\,arcseconds under photometric skies.

\subsection{Sky-Subtraction Procedure}
\begin{figure}
  \resizebox{\hsize}{!}{\includegraphics{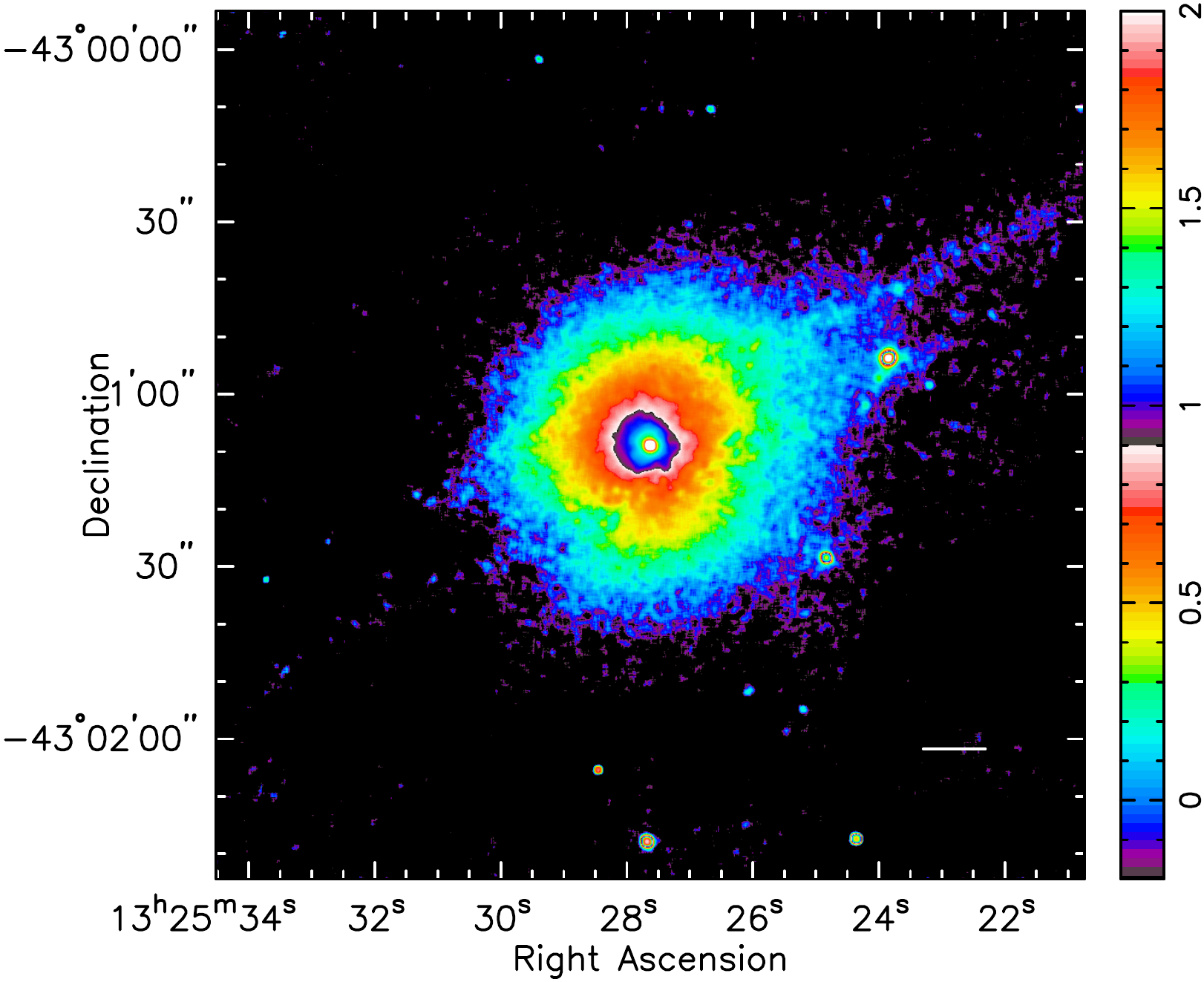}}
  \caption{Logarithmically-scaled 3.21~$\mu$m continuum emission from Cen\,A\@. The data have been
scaled by a factor of 1.042 (see text) and spatially smoothed by an $8\times8$ boxcar before log scaling.  The bar in the lower right corner illustrates 200\,pc at the distance to Cen\,A (3.8\,Mpc, \citet{Harris2010}). The wedge to the right shows relative surface brightness levels.}
  \label{Figure: Continuum}
\end{figure}

\begin{figure}
  \resizebox{\hsize}{!}{\includegraphics{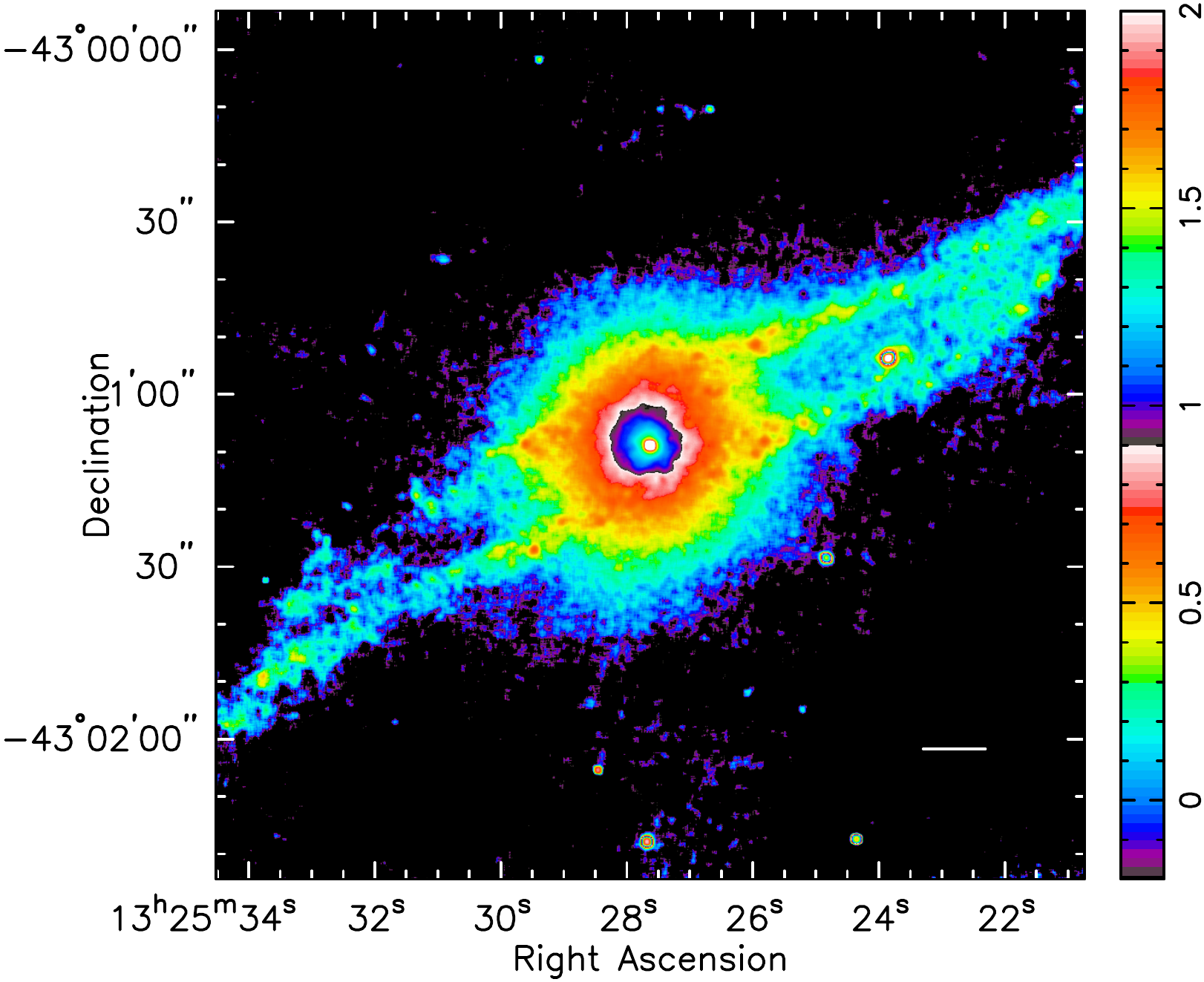}}
  \caption{Logarithmically-scaled 3.28~$\mu$m PAH feature+continuum emission from Cen\,A\@. The data have been
spatially smoothed by an $8\times8$ boxcar before log scaling.  The bar in the lower right corner illustrates 200\,pc at the distance to Cen\,A\@. The wedge to the right shows relative surface brightness levels.  Note that the scale is the same as in Fig.\ref{Figure: Continuum}.}
  \label{Figure: Feature+Continuum}
\end{figure}

\begin{figure}
  \resizebox{\hsize}{!}{\includegraphics{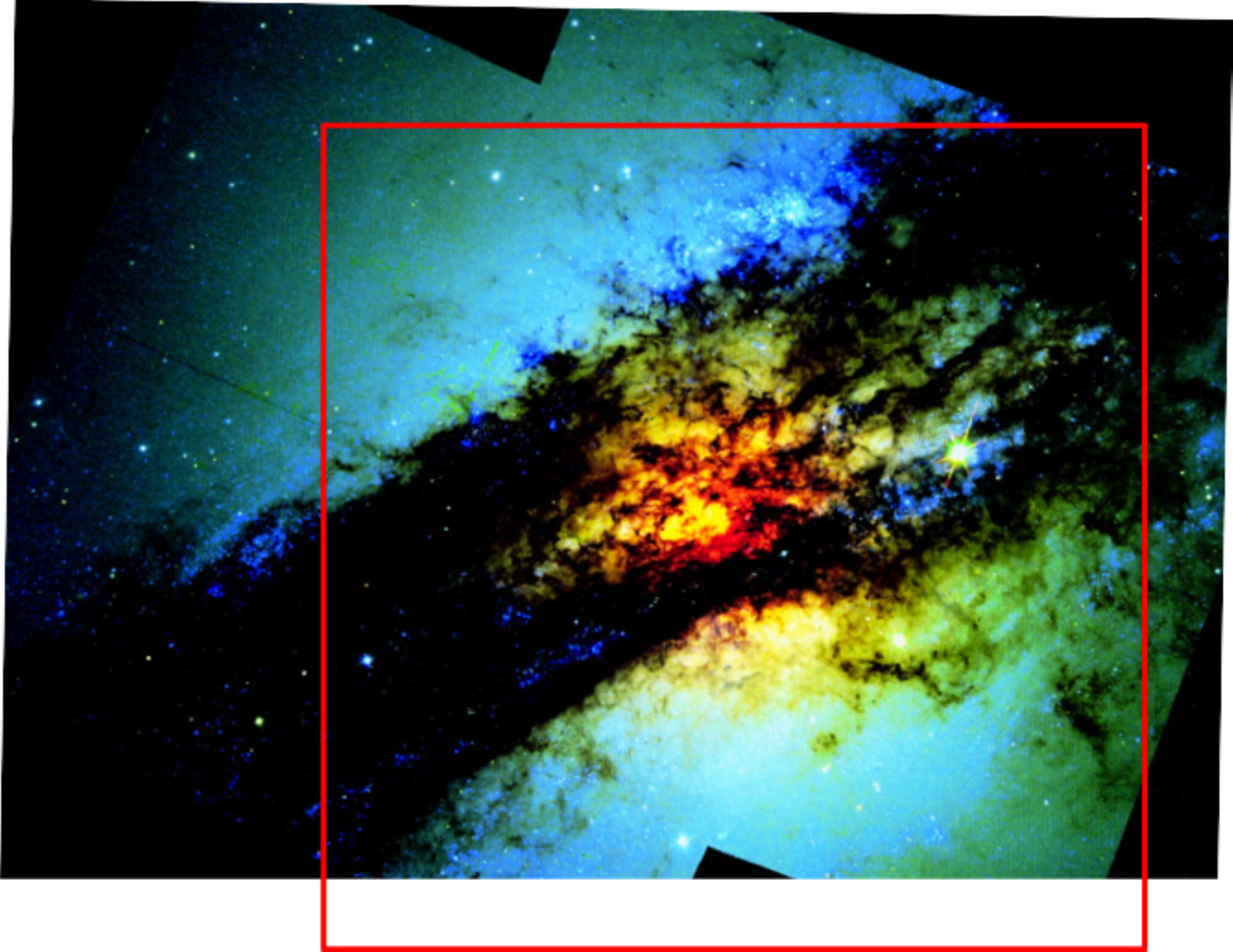}}
  \caption{Reproduction of the WFPC2 mosaic of Cen\,A \citep[Figure 4 of][]{Marconi2000} with the ISAAC field of view indicated with a red box.  For alignment with the ISAAC data it was necessary to apply a small (1\degr) rotation to the {\em HST\/} image.}
  \label{Figure: Marconi with FOV Overlay}
\end{figure}

\begin{figure}
  \resizebox{\hsize}{!}{\includegraphics{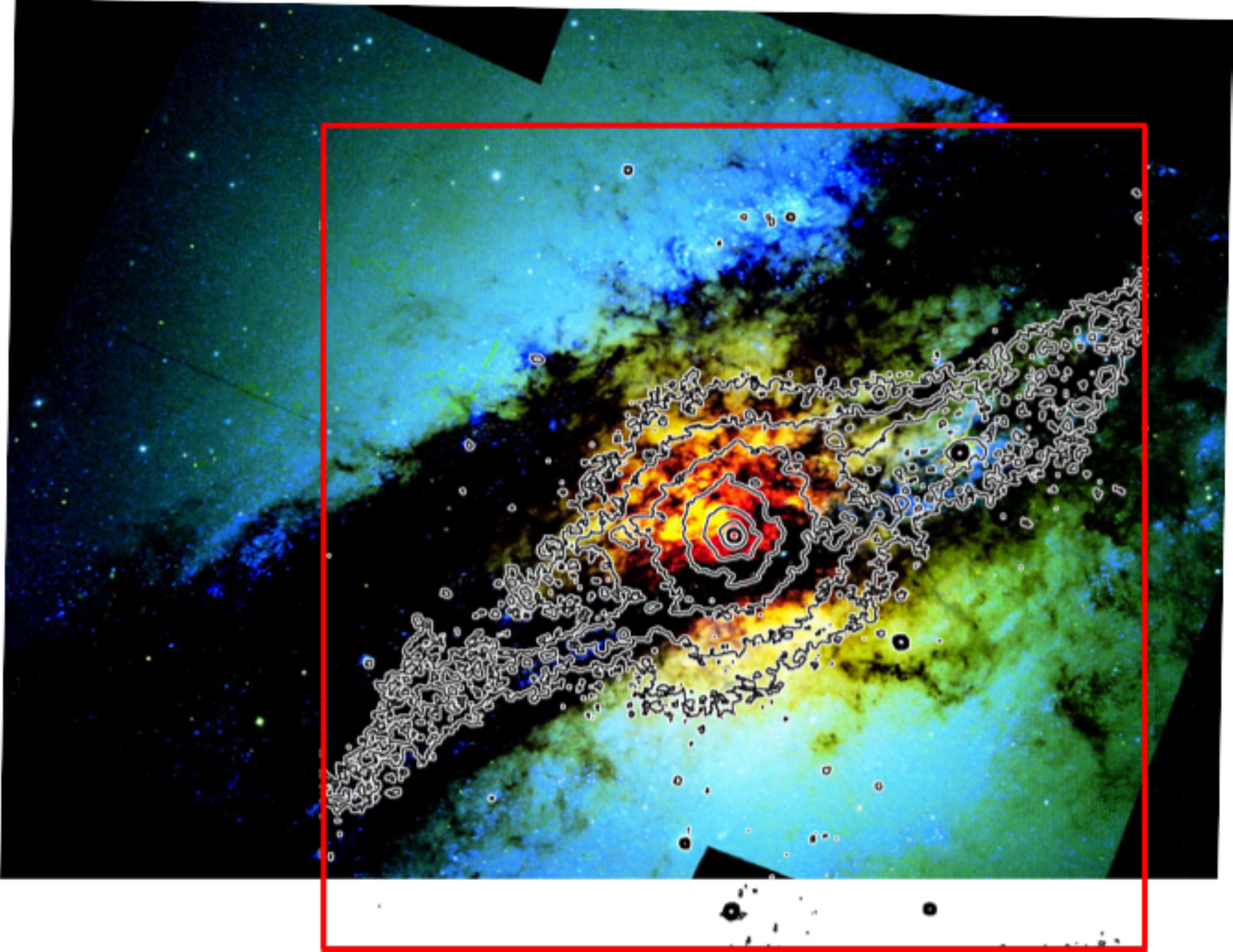}}
  \caption{The same as Fig.~\ref{Figure: Marconi with FOV Overlay} but with contours of the 3.28~$\mu$m PAH feature+continuum emission superposed.  The contours are based on the image shown in Fig.~\ref{Figure: Feature+Continuum} and have values of 0, 0.25, \ldots, 2.0 in the arbitrary units of the map.}
  \label{Figure: Marconi with L+C Contour Overlay}
\end{figure}

The data were reduced with IRAF\footnote{IRAF is distributed by the National Optical Astronomy Observatories,
    which are operated by the Association of Universities for Research
    in Astronomy, Inc., under cooperative agreement with the National
    Science Foundation.} using standard techniques as outlined in the {\em ISAAC Data Reduction Guide
1.5\/}\footnote{http://www.eso.org/sci/facilities/paranal/instruments/isaac/doc/drg/html/drg.html}.  However, during data reduction it was realized that the assumption of small source size adopted when planning the observations was unsubstantiated.  This meant that a procedure had to be adopted which went beyond the usual pair-wise subtraction of images to remove the sky emission.  The procedure adopted consisted of two passes.  The first pass involved simple pair-wise subtraction of images to remove sky, albeit imperfectly.  The resulting images were then co-aligned such that the very bright and well-defined nuclear peaks lined up.  A mask was then constructed from a spatially smoothed version of the co-addition of these images, such that the regions that contained source emission were zero and everywhere else was unity.  The final master mask was the superset of the two individual masks for the two filters.  That is, if there was emission seen in either individual mask it was set to zero in the master mask.  Knowing by how much each individual exposure had to be shifted to align the nuclear peak meant also knowing by how much the master mask had to be shifted in the opposite sense to mask the object in any given exposure.  In the second data reduction pass the sky for any given exposure was estimated from the median of the nearest exposures in which the nucleus was placed in the other quadrants, where in each case masking was applied as described.  The resulting sky-subtracted images were then co-aligned as in the first pass, by ensuring that the bright nuclear peaks were on top of each other.  For purposes of absolute coordinates we have fixed the coordinates of the nuclear peak to be that of the VLBI peak, 13$^\mathrm{h}$\,25$^\mathrm{m}$\,27\fs6152, $-$43\degr\,1\arcmin\,8.805\arcsec\ \citep{Ma1998}.  The resulting images from the NB\_3.21 filter (continuum) and NB\_3.28 filter (PAH+continuum) are shown in Figs.~\ref{Figure: Continuum} and \ref{Figure: Feature+Continuum}.  For direct comparison with optical observations of the same region we present in Figs.~\ref{Figure: Marconi with FOV Overlay} and \ref{Figure: Marconi with L+C Contour Overlay} a reproduction of Figure 4 of \cite{Marconi2000}.


\subsection{Continuum-Subtraction and Resulting Images}
\begin{figure}
\centering
{\includegraphics[width=8cm]{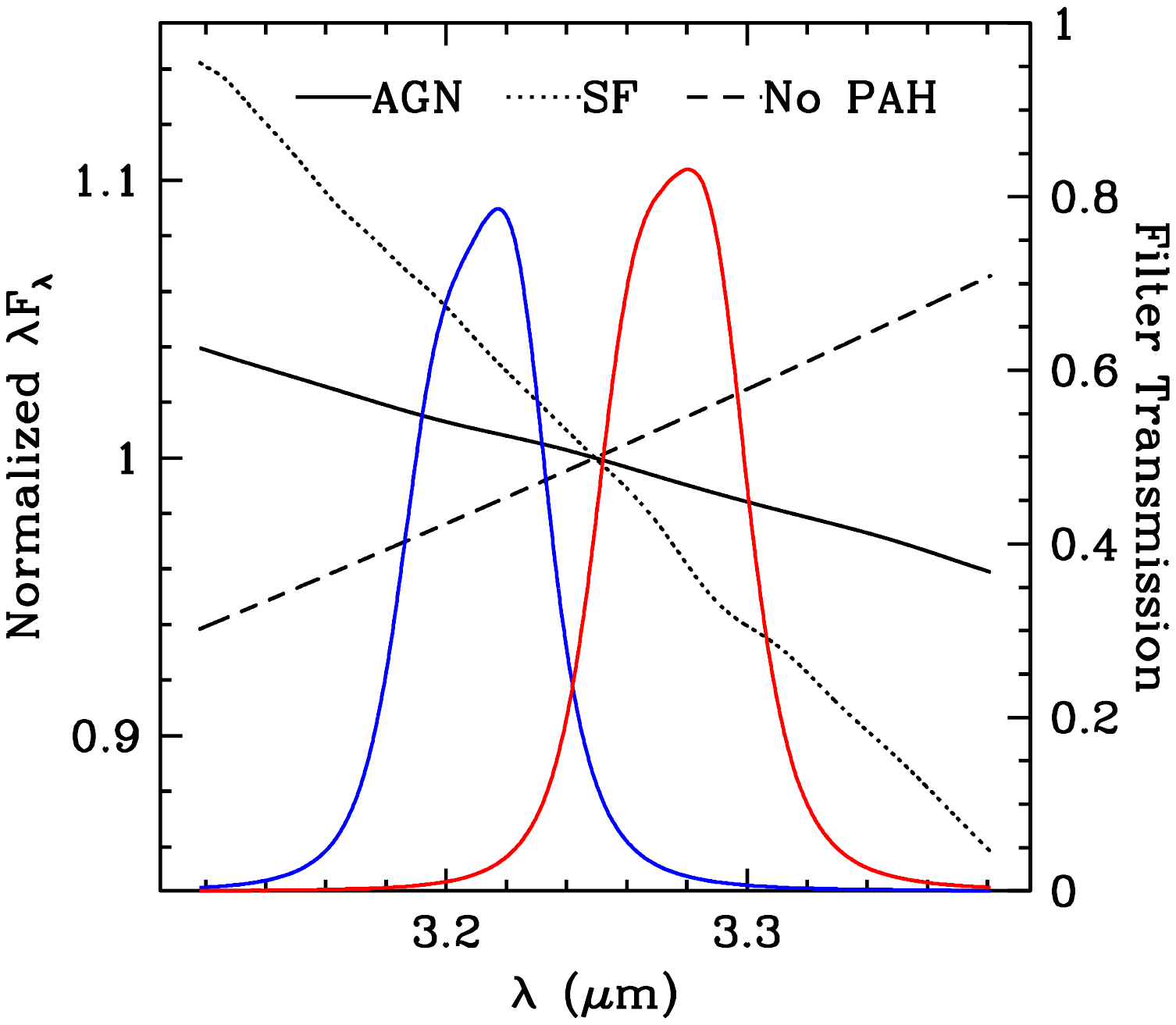}}
  \caption{Illustration of the differences of continuum slope as a function of the dominant contributor to the continuum.  The solid curve represents the AGN-dominated case \citep{Netzer2007}, while the dotted line shows our SB99 model results consistent with star formation \citep{Dale2005}.  The dashed line indicates the continuum slope that would have to be present to result in no detected PAH emission at the position of the nucleus (see text). The continua have been normalized to unity at $\lambda = 3.25~\mu$m.  For reference the two ISAAC filter curves are shown.}
  \label{Figure: ContinuumComparison}
\end{figure}

As preparation for continuum subtraction of the NB\_3.28 data, we modeled the continuum slope as one consistent with that of the range of nearby galaxies represented by the SINGS sample \citep{Kennicutt2003} or with an AGN-dominated spectrum.  In analysing the 1--850~$\mu$m SEDs of the SINGS galaxies \citet{Dale2005} find that they can be well fit by a combination of a dust-only component and a stellar component represented by a 900\,Myr continuous star formation with solar metallicity and Salpeter IMF (see their Figs.~3--10).  Over the wavelength range of our observations only the latter component contributes.  We have thus used Starburst99 to produce such a continuum spectrum (Fig.~\ref{Figure: ContinuumComparison}) and have determined the relative flux that would be observed with the two ISAAC filters if the continuum had that shape.  We find that in such a case scaling the flux in the NB\_3.21 filter image by a factor of 1.099 before subtracting from the NB\_3.28 image would remove the continuum from that image.

To model an AGN-dominated continuum spectrum  we relied on the result of \citet{Netzer2007} who have analysed the IR SEDs of QSOs in the QUEST sample \citep{Schweitzer2006}.  In their work they isolate a ``pure'' AGN continuum by subtracting a starburst template from the mean SEDs of their sample \citep[for details see][]{Netzer2007}.  We have fit their tabulated AGN continuum over the wavelength range of our observations using a simple polynomial (Fig.~\ref{Figure: ContinuumComparison}), and find that for this case scaling the flux in the NB\_3.21 filter by a factor of 1.042 before subtracting it from the flux in the NB\_3.28 filter would result in the latter being fully continuum-subtracted\footnote{Note that despite the fact that the model continua are both blue over this wavelength range, the derived continuum-correction factors are both larger than unity owing to the fact that the throughput of the NB\_3.28 filter is larger than that of the NB\_3.21 filter.}.

\begin{figure}
  \resizebox{\hsize}{!}{\includegraphics{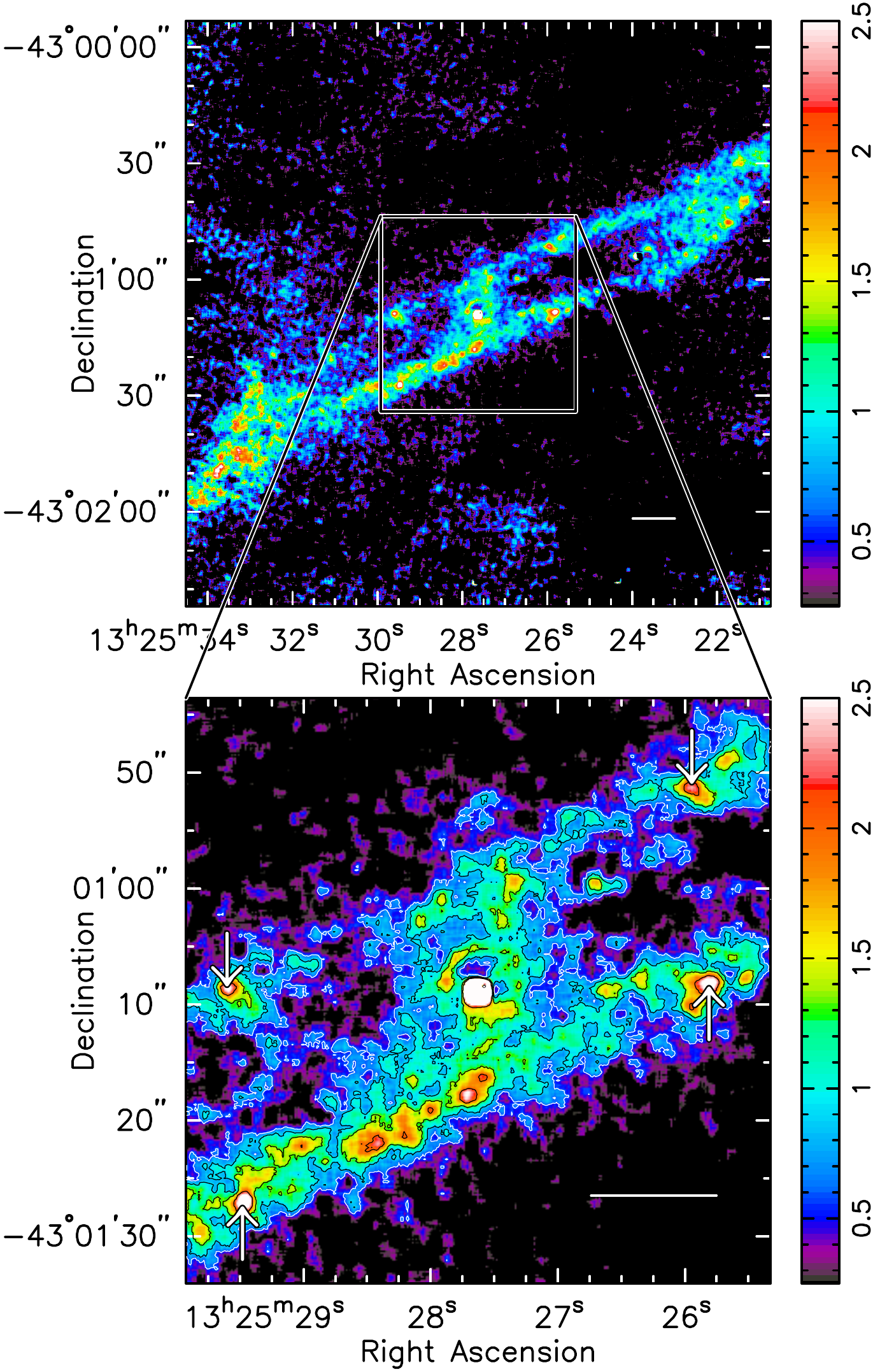}}
  \caption{Linearly-scaled 3.3~$\mu$m continuum-free PAH feature emission from Cen\,A, derived from the subtraction of the scaled NB\_3.21 data from the NB\_3.28 data (see text) and spatially smoothed by an $8\times8$ boxcar.  The bar in the lower right corner of each panel illustrates 200\,pc at the distance to Cen\,A\@. The wedges to the right show relative surface brightness levels.  The bottom panel shows a blowup of the central region to better illustrate details (including four prominent spots indicated with arrows, the nearly circular clumpy ring surrounding the nucleus (see also the blow-up of this ring in Fig.~\ref{Figure: ISAAC PAH/Paschen alpha}), and the faint arc-like feature north of the nucleus at about ($\alpha_{2000}, \delta_{2000} = 13^h\,25^m\,27\fs 6, -43\degr\,00^\prime \,52 ^{\prime\prime}$) described in the text.  Logarithmically-spaced contours are drawn at the levels 0.60, 0.90, 1.34, and 2.00 in the arbitrary units of the map.}
  \label{Figure: ISAAC PAH}
\end{figure}

Scaling our observed continuum by the star formation factor of 1.099 resulted in a continuum level near the nucleus which was higher than the PAH feature+continuum flux measured in the same region by the NB\_3.28 observations.  Thus the central regions are better represented by an AGN-dominated continuum slope.  Since the observed continuum levels in the regions which lie further from the nucleus are inherently much lower than those of the PAH+continuum image there is little difference in adopting one continuum scaling versus the other.  Hence, for the entire field of view we correct the flux in the NB\_3.21 image by a factor of 1.042 before subtracting it from the NB\_3.28 image to produce a continuum-free PAH feature image.  The resulting continuum-free 3.3~$\mu$m PAH feature image is presented in Fig.~\ref{Figure: ISAAC PAH}.  The emission at the position of the nucleus itself is consistent with a point source in the NB\_3.21 and NB\_3.28 observations, as well as in the resulting continuum-free PAH image.

\begin{figure}
  \resizebox{\hsize}{!}{\includegraphics{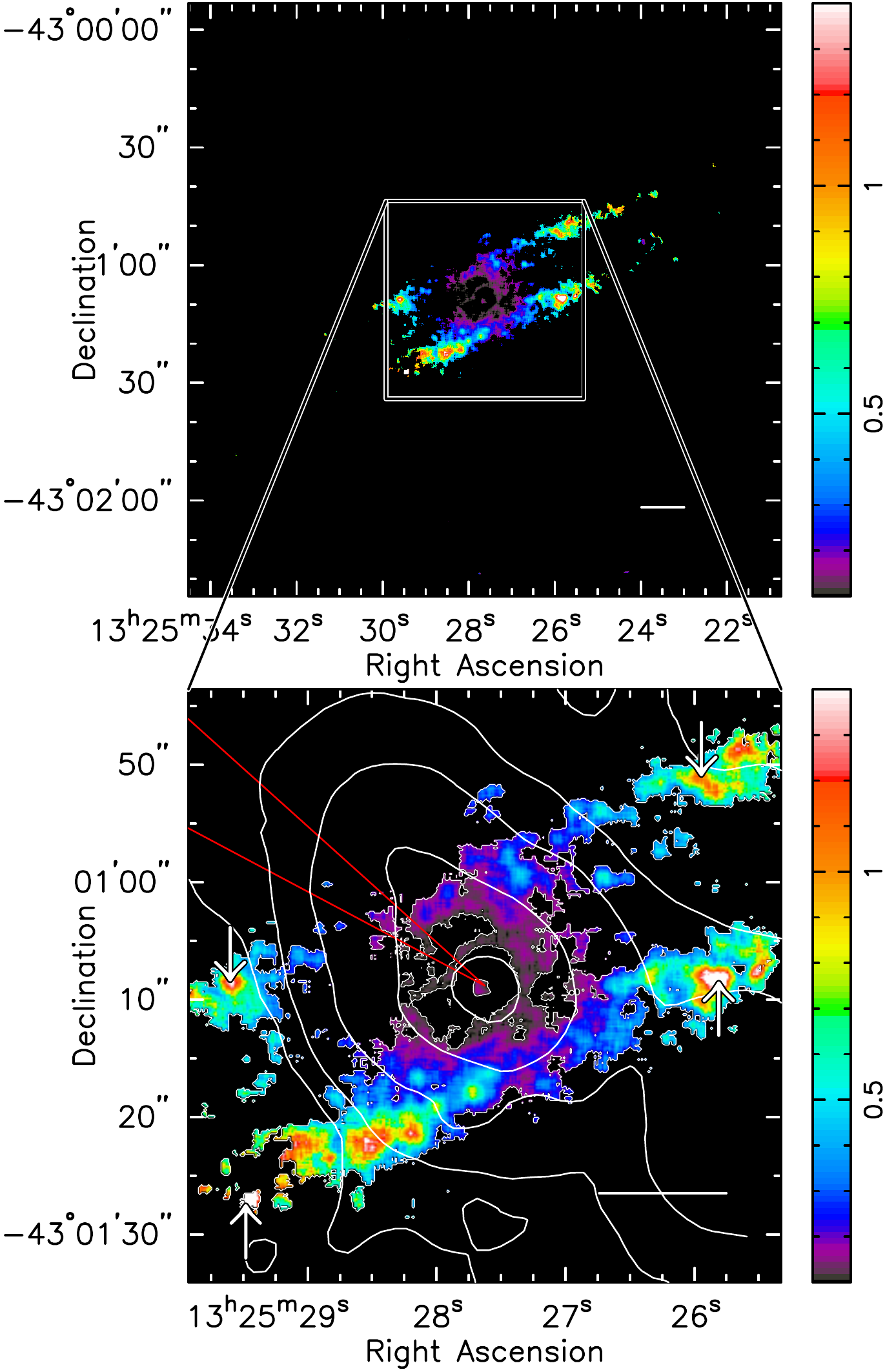}}
  \caption{3.3~$\mu$m PAH feature to continuum ratio emission from Cen\,A, derived from the images shown in Figs.~\ref{Figure: Continuum} and \ref{Figure: ISAAC PAH}.  The bar in the lower right corner of each panel illustrates 200\,pc at the distance to Cen\,A\@. The wedges to the right indicate the values of the ratio.  The bottom panel shows a blowup of the central region to better illustrate details.  The arrows in that panel are at the positions of the bright spots as defined in the PAH feature map (Fig.\ref{Figure: ISAAC PAH}) to facilitate comparison with that figure.  We plot a contour at a feature to continuum value of 0.1 to make the lowest levels more visible against the black background.  On top of the image we show logarithmically-spaced contours of the mid-IR [\ion{O}{iv}] emission \citep{Quillen2008} indicating the Narrow Line Region (see text) and two red lines indicative of the position angle of the radio/X-ray jet \citep{Burns1983,Kraft2000}.}
  \label{Figure: ISAAC PAH/Continuum}
\end{figure}

We have used the continuum-subtracted PAH feature map (Fig.~\ref{Figure: ISAAC PAH}) together with the pure continuum map (Fig.~\ref{Figure: Continuum}) to derive a feature to continuum ratio map.  During the division of these two maps only those pixels which were at least 3$\,\sigma$ from their respective mean background levels were considered.  The resulting 3.3~$\mu$m PAH feature-to-continuum map is shown in Fig.~\ref{Figure: ISAAC PAH/Continuum}.

\section{Results}
\subsection{Large Scale Features}
\label{Results: Large Scale Features}
The continuum emission presented in Fig.~\ref{Figure: Continuum} is strongly peaked towards the position of the nucleus, with faint traces of emission to the northwest and southeast.  As well there are clear indications of extinction due to the well-known dust lanes in Cen\,A\@.  The PAH feature+continuum emission map (Fig.~\ref{Figure: Feature+Continuum}), while also clearly peaked towards the nucleus, shows much more prominent emission towards the northwest and southeast with a parallelogram-shaped morphology.  The regions in the continuum image that display evidence for the extinction are completely filled in in this image.  These clear differences in morphology between the observations made with the NB\_3.21 and NB\_3.28 filters are strongly indicated in both the continuum-free PAH feature emission map (Fig.~\ref{Figure: ISAAC PAH}) as well as the PAH feature to continuum ratio map (Fig.~\ref{Figure: ISAAC PAH/Continuum}).

Figure~\ref{Figure: ISAAC PAH} reveals that it is the PAH feature emission that better traces out a symmetric parallelogram centered on the nucleus than does the 3.3~$\mu$m continuum emission.  This morphology was already seen in a number of different observations, covering a wide range of wavelengths (Table\ref{t1}).

\begin{table}[]
\caption{\label{t1}Other observations revealing the parallelogram structure in Cen\,A.}
\centering
\begin{tabular}{cccc}
\hline\hline
Wavelength/Line & Telescope/Instrument & References & Notes \\
\hline
H$\alpha$&AAT/TAURUS&1\\
1.25$\,\mu$m, 1.65$\,\mu$m, 2.2$\,\mu$m & NTT/SOFI & 2 & \tablefootmark{a} \\
3.6$\,\mu$m, 4.5$\,\mu$m, 5.8$\,\mu$m, 8.0$\,\mu$m &{\em Spitzer}/IRAC & 3, 4 \\
7$\,\mu$m, 15$\,\mu$m & {\em ISO}/ISOCAM & 5 \\
450$\,\mu$m, 850$\,\mu$m & JCMT/SCUBA & 5, 6 \\
1.3$\,$mm$^{12}$CO(2-1) & SMA & 7& \tablefootmark{b} \\
1.3$\,$mm$^{12}$CO(2-1) & ALMA & 8 \\
21\,cm \ion{H}{i}&ATCA&9\\
\hline
\end{tabular}
\tablefoot{
\tablefoottext{a} after extinction correction and ellipse subtraction;
\tablefoottext{b} central part of the structure only
}
\tablebib{
(1) \citet{Bland1987};
(2) \citet[][private communication]{Kainulainen2009};
(3) \citet{Quillen2006};
(4) \citet{Brookes2008};
(5)~\citet{Mirabel1999};
(6) \citet{Leeuw2002};
(7) \citet{Espada2009};
(8) ESO Photo Release eso1222, http://www.eso.org/public/news/eso1222/;
(9) \citet{Struve2010}.
}
\end{table}%

Finally, we note that the ISAAC data are the first to clearly show that the parallelogram feature is traced by continuum-subtracted PAH feature emission\footnote{\citet{Quillen2008} present {\em Spitzer Space Telescope\/} maps of the 7.7~$\mu$m, 8.6~$\mu$m, and 11.3~$\mu$m PAH features (their Figs.~13 and 3) in the central region of Cen\,A, though those maps are not continuum-subtracted.}.  This PAH feature emission is seen to consist of a number of knots throughout the parallelogram structure, but most notably in the inner regions (see the blowup in Fig.~\ref{Figure: ISAAC PAH}) and in the northwestern and southeastern ends.

\subsection{The Nucleus and Surrounding Regions}
\label{Results: Central Regions} 
In the central regions of Cen\,A (blowups in Figs.~\ref{Figure: ISAAC PAH} and \ref{Figure: ISAAC PAH/Continuum}) one sees a wealth of structure.  First, there are 4 bright knots as indicated with arrows in those figures.  These spots are the same as have been seen at 8 and 24~$\mu$m by \citet{Quillen2006Shell}, and are amongst a number of features which define the knotty ``sides" of the parallelogram structure.  The southeastern ``arm" shows more bright knots than the other parts of that structure.  Given the general radial falloff of the continuum emission, these knotty structures represent regions of relatively high feature-to-continuum ratio (see Section \ref{Discussion: Central Regions: The Four Knots} for discussion of the brightest four knots).  In addition, PAH feature emission is seen between these straight lines and the bright nucleus itself in a roughly north-south bar-like structure, which appears to end at a nearly complete, nearly circular clumpy ring of PAH feature emission about 6.5\arcsec\ (120\,pc) in diameter surrounding the nucleus (see also the blow-up of this ring in Fig.~\ref{Figure: ISAAC PAH/Paschen alpha}).  Unlike at the positions of the knots in the linear features, the locations of the clumps in the ring show only slight rises in the PAH feature-to-continuum ratio while the ring as a whole is visible in Fig.~\ref{Figure: ISAAC PAH/Continuum}.  

Interior to this ring, spatially unresolved nuclear PAH feature emission is clearly detected, despite the potentially hostile environment of the active nucleus.  Indeed it is impossible to scale the continuum emission prior to subtraction from the feature+continuum emission in such a way that the nuclear PAH feature disappears without strongly over-subtracting the continuum at all other locations with the field of view.  Moreover, if one considers only the nucleus itself, its SED would have to have a slope as indicated by the dashed line in Fig.~\ref{Figure: ContinuumComparison} for the PAH emission to vanish there.  The slope of that SED is physically unrealistic.  Thus, the nucleus does show a very low, but non-zero, feature-to-continuum emission ratio, perhaps explaining why the PAH feature emission has gone undetected to date \citep[e.g.][]{Laurent2000}.

This is not to suggest that the nucleus has no influence at all on the PAH emission.  In Fig.~\ref{Figure: ISAAC PAH/Continuum} one clearly sees the feature to continuum decreasing roughly radially towards the nucleus.  Indeed, the contours of [\ion{O}{iv}] could serve almost as well as contours of the  PAH feature to continuum ratio.  Thus, assuming that the [\ion{O}{iv}] does trace the Narrow Line Region rather than local shocks\footnote{\citet{Krajnovic2007} cite evidence for the radial increase in the importance of shocks though their analysis is restricted to the central few arcseconds.}, the nucleus is seen to have a marked effect on the efficiency of the PAH emission in its vicinity.  There is no obvious impact on the PAH emission stemming from the radio/X-ray jet (see Fig.~\ref{Figure: ISAAC PAH/Continuum} where lines indicate the position of the jet).

Finally, there is a faint, arc-like feature $\sim$15\arcsec\ to the north of the nucleus seen in the PAH feature image and marginally in the feature-to-continuum image.  There is no detected counterpart of this feature to the south of the nucleus.

\section{Discussion}
\label{Discussion}
\subsection{Large Scale Features}
\label{Discussion: Large Scale Features}
The fact that the 3.3~$\mu$m PAH feature traces out the large, symmetric parallelogram may not be surprising in light of the fact that it is better traced by the longer wavelength (5.8~$\mu$m and 8.0~$\mu$m) IRAC imaging than at the shorter IRAC bands \citep[3.6~$\mu$m and 4.5~$\mu$m;][]{Quillen2006}.   The longer wavelength bands are more dominated by a dust component and therefore less influenced by (diffuse) starlight than is the case at the shorter wavelength bands.  However, what is less anticipated is the fact that the emission from the parallelogram is so faint relative to the nucleus in the 3.3~$\mu$m continuum.  The feature is seen at 3.6~$\mu$m and at $J$ (1.247~$\mu$m) and K$_{\mathrm s}$ (2.162~$\mu$m), once the veil of $\sim$0.4$\,$magnitudes of extinction is removed \citep[][their Figs.~1 and 2, also visible at $H$  (1.653~$\mu$m), Kainulainen, private communication]{Kainulainen2009}.  In the $J$ image the large numbers of fairly bright sources seen in the extended structure are about a factor of 300 fainter than the nuclear peak.  The brightness ratio in the case of the 3.6~$\mu$m imaging is not easy to infer from Fig.~1 of \citet{Quillen2006} but is likely not substantially lower than this.  This would be consistent with results of \citet{Dale2005} and of \citet{Netzer2007} that together would imply similar ratios of the nuclear brightness to that in the larger scale structure.  The dynamic range in our 3.3~$\mu$m continuum image (Fig.~\ref{Figure: Continuum}) is about a factor of 125\@.  Thus, it may be the case that our observations are not deep enough to have detected the continuum at this wavelength.  Nevertheless, the fact that we have made a clear detection of the PAH feature emission, while at the same time not having detected the continuum emission means that at this wavelength the emission from the parallelogram is clearly dominated by dust emission.  This is in contrast to the case at 3.6~$\mu$m.

\subsection{Inner Kiloparsec}
\label{Discussion: Inner Kiloparsec}
\begin{figure}
  \resizebox{\hsize}{!}{\includegraphics{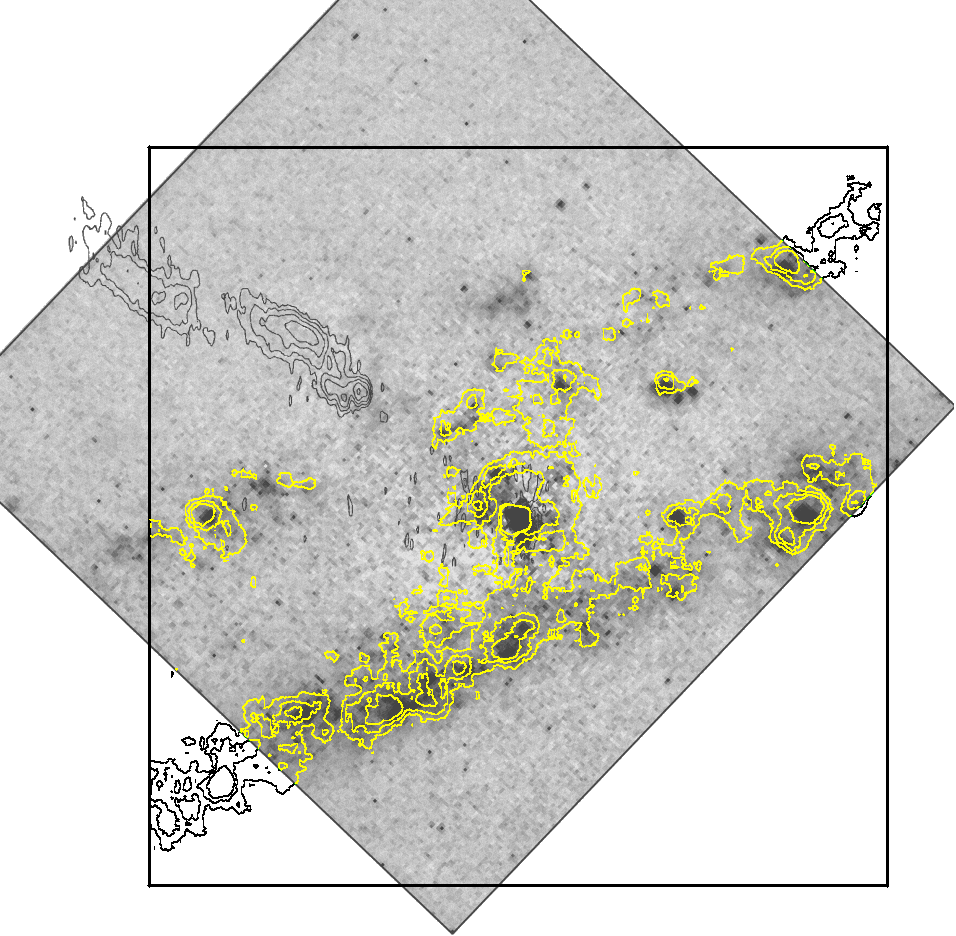}}
  \caption{A comparison of the 3.3~$\mu$m PAH feature emission (yellow/black contours), Pa$\alpha$ emission (underlying grayscale image), and 6\,cm radio continuum emission \citep[light black contours, the latter two taken from Fig.~10 of][]{Marconi2000}.}
  \label{Figure: ISAAC PAH / Paa Comparison}
\end{figure}

That the 3.3~$\mu$m PAH feature emission is related to star formation is demonstrated in Fig.~\ref{Figure: ISAAC PAH / Paa Comparison}.  In this figure we show as a background image a colorized version of Fig.~10 (right) of \citet{Marconi2000}, illustrating {\em HST\/} Pa$\alpha$ data with {\em VLA\/} 6\,cm radio continuum contours.  Overlaid are dark contours of PAH feature emission.  The dark box surrounding the contours represents the same field of view as is illustrated in the bottom panels of Figs.~\ref{Figure: ISAAC PAH} and \ref{Figure: ISAAC PAH/Continuum}.  Alignment between these two images was achieved in a two step manner.  First, we aligned the 6\,cm radio continuum contours from Fig.~3 (bottom) of \citet{Burns1983} with those in Fig.~10 (right) of \citet{Marconi2000}.  This allowed us to clearly identify the nuclear position in the latter image.  The second step was to then place our PAH image on top of the nuclear position defined by that of \citet{Burns1983}, being sure to match the spatial scales.

\begin{figure}
 \centering
  {\includegraphics[width=8cm]{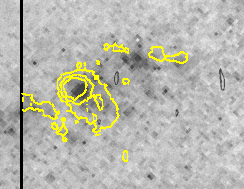}}
  \caption{A blowup of the northeast portion of the Pa$\alpha$ and PAH feature emission shown in Fig.~\ref{Figure: ISAAC PAH / Paa Comparison}, illustrating a spatial anti-corelation between the two emission features.}
  \label{Figure: ISAAC PAH / Paa Comparison Blow-up}
\end{figure}

We see in this figure a generally very good agreement in the overall distribution of the PAH feature emission with that of the Pa$\alpha$, but with some small offsets and other discrepancies clearly visible.  For example, while the PAH emission is seen to align quite well with the features in the brighter SE part of the parallelogram, the PAH emission at the positions of the three bright knots (as indicated by arrows in Fig.~\ref{Figure: ISAAC PAH}, the fourth knot lying outside the field of view of the Pa$\alpha$ data) is seen to be displaced from the position of the Pa$\alpha$ peaks.  Another clear mis-alignment example is the case of the Pa$\alpha$ patch to the northwest of the northeastern bright PAH/Pa$\alpha$ knot (Fig.\ref{Figure: ISAAC PAH / Paa Comparison Blow-up}).  In that case the PAH emission appears to lie on either side of the Pa$\alpha$ but not at its position.  (See also Fig.~\ref{Figure: ISAAC PAH/Paschen alpha} for similar offsets between Pa$\alpha$ and PAH in the very center of Cen\,A).  Any reasonable extinction law would indicate that the extinction for Pa$\alpha$ is higher than at 3.3~$\mu$m, which implies that one cannot account for such spatial distribution differences with differential extinction. 

\begin{figure}
  \resizebox{\hsize}{!}{\includegraphics{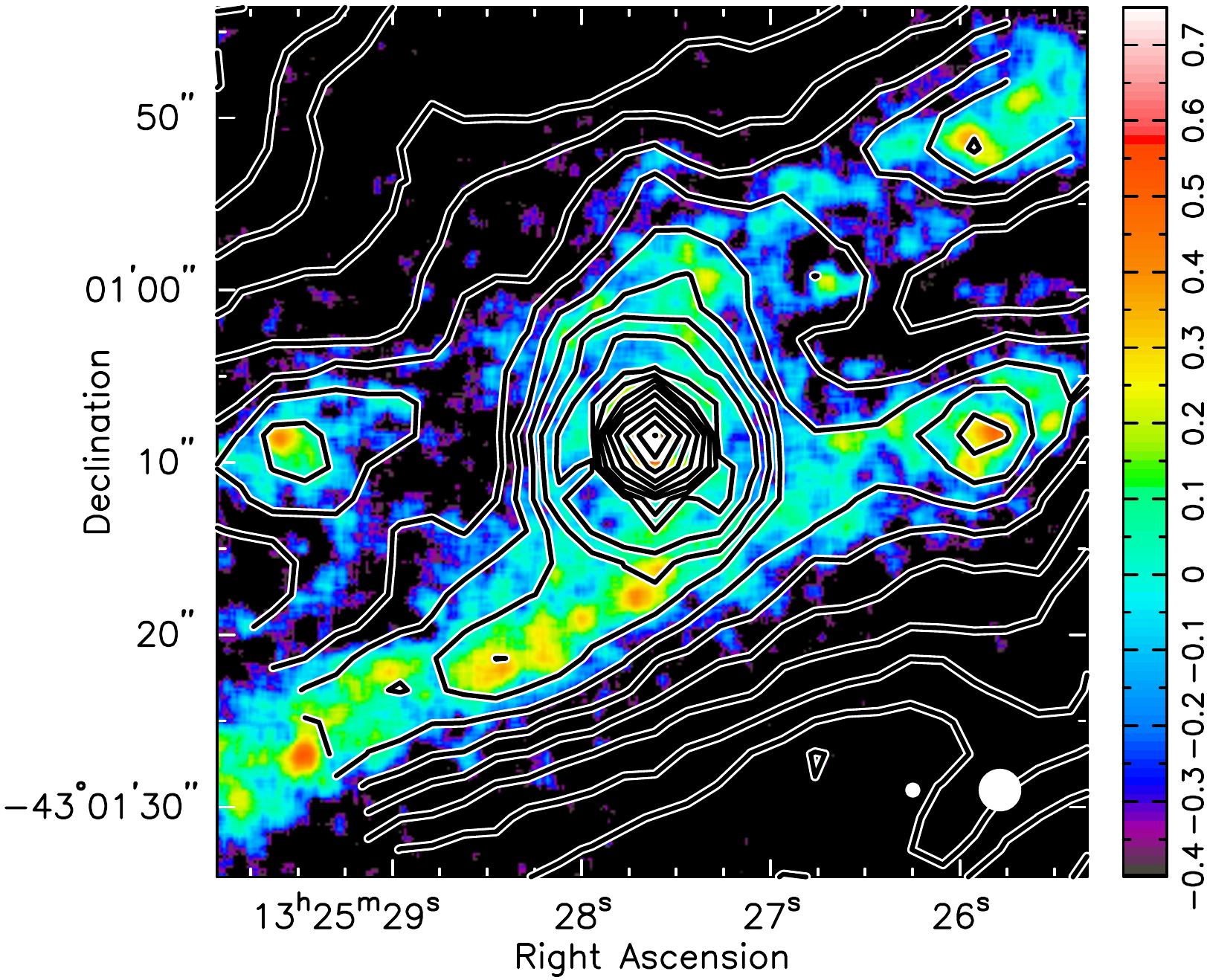}}
  \caption{3.3~$\mu$m PAH feature emission (image) with contours of [\ion{Ne}{ii}] emission from \cite{Quillen2008}.  The white circles in the lower right represent the spatial resolution of the ISAAC observations (left) and the {\em Spitzer\/} observations (right).}
  \label{Figure: ISAAC PAH / [NeII] Comparison}
\end{figure}

As an alternate tracer of star formation we present in Fig.~\ref{Figure: ISAAC PAH / [NeII] Comparison} the 12.8~$\mu$m [\ion{Ne}{ii}] emission from \citet{Quillen2008} as contours on top of the ISAAC 3.3~$\mu$m PAH feature emission from the present work.  The [\ion{Ne}{ii}] observations also show a relatively good agreement with the PAH peaks, though given the resolution disparity of a factor of $\sim$3.7 between the two data sets we cannot conclusively rule out that there are spatial offsets between the two emission features. 

\subsection{The 500\,pc Shell}
\label{Discussion: The Shell}
\citet{Quillen2006Shell} report on the discovery of a shell of emission some 30\arcsec\ in radius from the nucleus of Cen\,A in {\em Spitzer\/} observations at 4.5, 5.8, 8, and 24~$\mu$m \citep[Fig.~1 of][]{Quillen2006Shell}.  The nature of the shell is unknown at present, at least in part owing to a current lack of velocity information.  Quillen et al.~present plausibility arguments that suggest that a moderate starburst a few thousand solar masses interior to the shell could have led to its formation.  Further, based on size arguments and the perceived requirement for a gap in the dust distribution \citep{Quillen2006}, they note that there could exist multiple circum-nuclear shells in Cen\,A\@.

The fact that the shell detected in the {\em Spitzer\/} observations is undetected in the current study is puzzling in light of our NGC\,253 3.3~$\mu$m imaging results \citep{Tacconi-Garman2005}.  In that work we found the first evidence for PAH molecules in a starburst-driven galactic superwind.  The Cen\,A shell is similar to though smaller in scale than the one seen in NGC\,253\@.  Ratios of PAH emission at various wavelengths (especially to emission at 3.3~$\mu$m) can be used as indicators of grain size, grain photoionization state, etc.~\citep[e.g.][]{Bauschlicher2009}.  
Our non-detection at 3\,microns could indicate that the Cen\,A shell contains preferentially large grains \citep{Bregman1994}, perhaps as a result of smaller grains having been destroyed \citep{Smith2007}, though this seems implausible, especially in light of our detection of 3.3~$\mu$m PAH emission in the nucleus itself.  An alternative explanation could be that the PAH grains in the shell are preferentially ionized.  This would tend to strengthen C-C stretching mode transitions (6.2, 7.7\footnote{The 7.7~$\mu$m transition is a blend of C-C stretching and C-H in-plane bending \citep{Bauschlicher2008,Bauschlicher2009}.}, and 8.6~$\mu$m) relative to the C-H stretching mode at  3.3~$\mu$m \citep[][and references therein]{vanDiedenhoven2004}.  Further, deeper observations at 3\,microns would be required to better understand the composition and physical state of the shell, perhaps shedding light on its origin.

While there is no current detection of the 3.3~$\mu$m PAH feature emission counterpart to the shell at 30\arcsec\ radius, the detection of the faint, arc-like feature $\sim$15\arcsec\ to the north perhaps suggests that there have indeed been episodic events that have resulted in such shells.  In the Pa$\alpha$ data of \citet{Marconi2000} there is also a suggestion of an arc roughly 20\arcsec\ north of the nucleus (see Fig.~\ref{Figure: ISAAC PAH / Paa Comparison}), lending further support for this.  Although the {\em Spitzer\/} imaging shows evidence for a southern counterpart for the northern loop of the shell, neither the Pa$\alpha$ arc at 20\arcsec\ nor the 3.3~$\mu$m are at 15\arcsec\ has a counterpart to the south of the nucleus.

\subsubsection{The Four Knots}
\label{Discussion: Central Regions: The Four Knots}
The four knots that are clearly evident in 3.3~$\mu$m PAH feature emission (Figs.~\ref{Figure: ISAAC PAH} and \ref{Figure: ISAAC PAH/Continuum}) are also present in numerous other line observations.  They are seen in {\em Spitzer\/} observations of [\ion{S}{iii}], [\ion{Si}{ii}], [\ion{Fe}{iii}], [\ion{Ne}{ii}], H$_2$S(0), and to a lesser extent [\ion{Fe}{ii}] \citep{Quillen2008}.  In addition, three of the four spots are also evident in the Pa$\alpha$ imaging of \citet{Marconi2000}, with the last one falling outside their field of view.  The fact that these spots are evident in as many line and continuum observations as they are suggests that they cannot be due to a chance superposition of the warped disk and the shell, as posited by \citet{Quillen2006Shell}.  Rather, it appears more likely that these locations represent locations at which star formation is enhanced, as all of these lines can be found at or near sites of active star formation.  Indeed, it may be the case that an interaction between an expanding shell and the warped disk at those locations has resulted in an enhancement of molecular cloud compression and hence an increase in the local start formation rate.  Certainly knowledge of the kinematics of the shell would not only help in defining it as a true coherent structure as noted by \citet{Quillen2006Shell}, but would also yield clues as to what influence it could have when confronting gas in the warped disk.

Finally, we note that if these knots do represent sites of enhanced star formation then there must be differences in the physical conditions and/or grain population relative to those at similar sites in the starburst galaxies NGC\,253 and NGC\,1808.  In both those cases \citet{Tacconi-Garman2005} found a decrease in the PAH feature-to-continuum ratio at star formation sites, while in the present case the same ratio is seen to peak at the positions of the knots.  This implies that neither mechanical energy input into the local ISM nor photoionisation of the PAH nor photodissociation of the PAH, all of which could explain a drop in the efficiency of 3.3$\,\mu$m PAH emission \citep{Tacconi-Garman2005} can play dominant roles at the location of these knots.  It could though be the case that the stellar contribution to the continuum is lower at these sites than it is in NGC\,253 and NGC\,1808.  Spatially resolved PAH observations at longer wavelengths as well as consistent star formation age-dating observations of all of these systems would be invaluable to better understand the presence of feature-to-continuum variations at these sites.

\subsubsection{The S-Shape}
\label{Discussion: Central Regions: The S-Shape}
\begin{figure}
  \resizebox{\hsize}{!}{\includegraphics{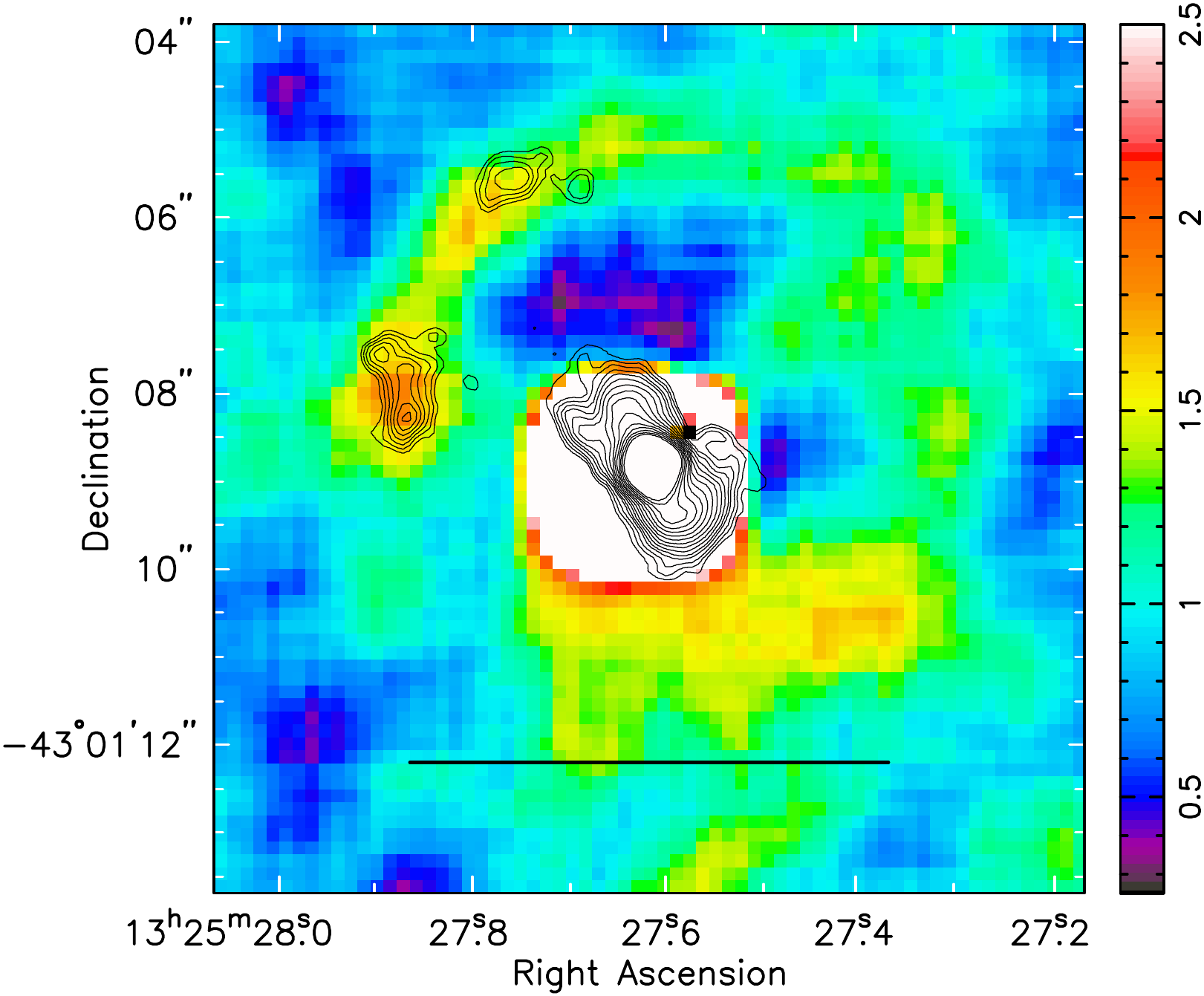}}
  \caption{3.3~$\mu$m PAH feature emission with contours of Paschen $\alpha$ emission \citep[][their Fig.~4]{Schreier1998}.  The bar below the nucleus represents 100\,pc at the distance to Cen\,A\@.}
  \label{Figure: ISAAC PAH/Paschen alpha}
\end{figure}

To explain the trapezoidal structure seen in the mid-infrared, \citet{Quillen2006} modeled the system using a warped thin disk.  In order to avoid the appearance in the model of an unobserved bright central bar-like feature, they posited a truncation of the dust distribution over the range $6\arcsec\la r\la 50\arcsec$.  They explored the possibility of mimicking a gap through decreasing the inclination of the innermost disk, but dismissed that as a viable alternative since the resulting model (their Fig.~10b) does not resemble their IRAC data as well as does a model with a gap.  Nevertheless, the nearly circular (or perhaps spiral) clumpy ring of PAH feature emission apparent in our observations (Fig.~\ref{Figure: ISAAC PAH} and blown up in Fig.~\ref{Figure: ISAAC PAH/Paschen alpha}) may be indicative of a more face-on structure in the very center of Cen\,A.  Note that this PAH feature bears further confirmation but that support for its existence is given by the rough spatial coincidence with observed Pa$\alpha$ emission \citep{Marconi2000}.

More recently \citet{Espada2009} have presented interferometric $^{12}$CO $J=2\rightarrow1$ observations of the inner 
$24\arcsec\times12\arcsec$ of Cen\,A\@.  While their data is broadly consistent with a morphology like that adopted by \citet{Quillen2006}, they do see departures from what such a model would predict in terms of both spatial distribution of gas and, importantly, the gas kinematics.  In addressing these differences they find that a combination of a warped disk model together with a weak central bar potential could qualitatively explain their CO observations.  In particular, such a model naturally reproduces the S-shape morphology seen in the interferometric CO data.  This morphology is also seen in the H$_2$S(0)$J=2$--0 line at 28.2~$\mu$m, but is not evident in the H$_2$S(3)$J=5$--3 line at 9.665~$\mu$m or the H$_2$S(5)$J=7$--5 line at 6.909~$\mu$m, as was already noted by \citet{Quillen2008}.  The fact that the H$_2$S(0) line emission shows a similar morphology with that of the CO leads Espada et al. to suggest that the warm molecular gas traces the locations of weak-bar-induced spiral arms, where the gas is heated by the resulting shocks and star formation in the arms.  As further support for this idea they claim that the data of \citet{Quillen2008} show that the star formation tracers of [\ion{Si}{ii}], [\ion{S}{iii}] and [\ion{Fe}{ii}] are seen to be brighter on the NW and SE sides of the inner parallelogram.  While some of the structure seen in those data may be related to the S-shaped morphology seen in the CO and H$_2$S(0) there are departures from this which weaken this interpretation.  For example, in both the [\ion{S}{iii}] and [\ion{Si}{ii}] maps \citep[Fig.~4 of][]{Quillen2008} the SW side is brighter than the NW arm.  This is also seen to be the case in the Pa$\alpha$ observations of \citet{Marconi2000} as well as in our own measurements (see Fig.~\ref{Figure: ISAAC PAH}).  The origin of this discrepancy remains unclear.

\section{Conclusions}
\label{Conclusions}
Based on our 3.3$\,\mu$m PAH imaging of Cen\,A we find that the nucleus is not devoid of PAH emission, but rather is very bright, despite the fact that the feature-to-continuum emission, which decreases towards the AGN, is lowest at that position.  This means that the PAH particles are not systematically destroyed in the nucleus as might be expected for such a harsh environment.  Perhaps this is due to a large quantity of dust having been deposited in and near the nucleus as a result of the same merging process believed to have caused the larger scale dust structures seen in this galaxy.  However, we note that this may not be the whole explanation, as similar results have been obtained in 3.3$\,\mu$m imaging of NGC\,1068 (Tacconi-Garman, in preparation).

In addition, the 3.3$\,\mu$m PAH feature emission generally traces the sites of star formation in Cen\,A, but in detail there are spatial offsets.  This is as was seen in the earlier study of the starburst galaxies NGC\,253 and NGC\,1808 \citep{Tacconi-Garman2005}.  However, unlike in those other cases the feature-to-continuum ratio is not seen to drop at the sites of star formation.  Instead the ratio peaks at those positions, most notably at the locations of the four bright knots discussed in Section \ref{Discussion: Central Regions: The Four Knots}.  This implies that neither mechanical energy input into the local ISM nor photoionisation of the PAH nor photodissociation of the PAH, all of which could explain a drop in the efficiency of 3.3$\,\mu$m PAH emission \citep{Tacconi-Garman2005} can play dominant roles at the location of these knots.  In addition, what remains unclear at present is the age of the star formation in the knots, especially when compared to those in NGC\,253 and NGC\,1808.

Finally, our data reveal possible evidence for a nearly face-on, circular or spiral, dust structure surrounding the nucleus.  Although this feature bears confirmation by additional observations, support for its existence is given by the rough spatial coincidence with observed Pa\,$\alpha$ emission \citep{Marconi2000}.

\begin{acknowledgements}
The authors would like to extend their thanks to the anonymous referee for comments which resulted in a clearer and stronger paper.  As well, the authors wish to express their gratitude to the dedicated staff on Paranal for their support of these Service Mode observations.  Finally, the authors wish to warmly thank Alice Quillen, Joel Green, and Jouni Kainulainen for making the {\em Spitzer\/} and SOFI data available to us, and to Ethan Schreier, Alessandro Capetti, and Alessandro Marconi for their help in locating the original of a previously published image.
\end{acknowledgements}
\bibliographystyle{aa} 
\bibliography{post-referee-1.bib}

\begin{thebibliography}{52}
\expandafter\ifx\csname natexlab\endcsname\relax\def\natexlab#1{#1}\fi

\bibitem[{{Allamandola} {et~al.}(1989){Allamandola}, {Tielens}, \&
  {Barker}}]{Allamandola1989}
{Allamandola}, L.~J., {Tielens}, A.~G.~G.~M., \& {Barker}, J.~R. 1989, \apjs,
  71, 733

\bibitem[{{Bauschlicher} {et~al.}(2009){Bauschlicher}, {Peeters}, \&
  {Allamandola}}]{Bauschlicher2009}
{Bauschlicher}, C.~W., {Peeters}, E., \& {Allamandola}, L.~J. 2009, \apj, 697,
  311

\bibitem[{{Bauschlicher} {et~al.}(2008){Bauschlicher}, {Peeters}, \&
  {Allamandola}}]{Bauschlicher2008}
{Bauschlicher}, Jr., C.~W., {Peeters}, E., \& {Allamandola}, L.~J. 2008, \apj,
  678, 316

\bibitem[{{Bland} {et~al.}(1987){Bland}, {Taylor}, \& {Atherton}}]{Bland1987}
{Bland}, J., {Taylor}, K., \& {Atherton}, P.~D. 1987, \mnras, 228, 595

\bibitem[{{Bolatto} {et~al.}(2007){Bolatto}, {Simon}, {Stanimirovi{\'c}}, {van
  Loon}, {Shah}, {Venn}, {Leroy}, {Sandstrom}, {Jackson}, {Israel}, {Li},
  {Staveley-Smith}, {Bot}, {Boulanger}, \& {Rubio}}]{Bolatto2007}
{Bolatto}, A.~D., {Simon}, J.~D., {Stanimirovi{\'c}}, S., {et~al.} 2007, \apj,
  655, 212

\bibitem[{{Boselli} {et~al.}(2004){Boselli}, {Lequeux}, \&
  {Gavazzi}}]{Boselli2004}
{Boselli}, A., {Lequeux}, J., \& {Gavazzi}, G. 2004, \aap, 428, 409

\bibitem[{{Bregman} {et~al.}(1994){Bregman}, {Larson}, {Rank}, \&
  {Temi}}]{Bregman1994}
{Bregman}, J., {Larson}, K., {Rank}, D., \& {Temi}, P. 1994, \apj, 423, 326

\bibitem[{{Brookes} {et~al.}(2008){Brookes}, {Keene}, {Quillen},
  {Charmandaris}, {Lawrence}, {Stern}, \& {Werner}}]{Brookes2008}
{Brookes}, M.~H., {Keene}, J., {Quillen}, A.~C., {et~al.} 2008, in Astronomical
  Society of the Pacific Conference Series, Vol. 381, Infrared Diagnostics of
  Galaxy Evolution, ed. {R.-R.~Chary, H.~I.~Teplitz, \& K.~Sheth}, 405--+

\bibitem[{{Burns} {et~al.}(1983){Burns}, {Feigelson}, \&
  {Schreier}}]{Burns1983}
{Burns}, J.~O., {Feigelson}, E.~D., \& {Schreier}, E.~J. 1983, \apj, 273, 128

\bibitem[{{Calzetti}(2010)}]{Calzetti2010}
{Calzetti}, D. 2010, ArXiv e-prints

\bibitem[{{Dale} {et~al.}(2005){Dale}, {Bendo}, {Engelbracht}, {Gordon},
  {Regan}, {Armus}, {Cannon}, {Calzetti}, {Draine}, {Helou}, {Joseph},
  {Kennicutt}, {Li}, {Murphy}, {Roussel}, {Walter}, {Hanson}, {Hollenbach},
  {Jarrett}, {Kewley}, {Lamanna}, {Leitherer}, {Meyer}, {Rieke}, {Rieke},
  {Sheth}, {Smith}, \& {Thornley}}]{Dale2005}
{Dale}, D.~A., {Bendo}, G.~J., {Engelbracht}, C.~W., {et~al.} 2005, \apj, 633,
  857

\bibitem[{{Diamond-Stanic} \& {Rieke}(2010)}]{Diamond-Stanic2010}
{Diamond-Stanic}, A.~M. \& {Rieke}, G.~H. 2010, \apj, 724, 140

\bibitem[{{Draine} \& {Li}(2001)}]{Draine2001}
{Draine}, B.~T. \& {Li}, A. 2001, \apj, 551, 807

\bibitem[{{Engelbracht} {et~al.}(2005){Engelbracht}, {Gordon}, {Rieke},
  {Werner}, {Dale}, \& {Latter}}]{Engelbracht2005}
{Engelbracht}, C.~W., {Gordon}, K.~D., {Rieke}, G.~H., {et~al.} 2005, \apjl,
  628, L29

\bibitem[{{Espada} {et~al.}(2009){Espada}, {Matsushita}, {Peck}, {Henkel},
  {Iono}, {Israel}, {Muller}, {Petitpas}, {Pihlstr{\"o}m}, {Taylor}, \&
  {Dinh-V-Trung}}]{Espada2009}
{Espada}, D., {Matsushita}, S., {Peck}, A., {et~al.} 2009, \apj, 695, 116

\bibitem[{{Galliano} {et~al.}(2008){Galliano}, {Madden}, {Tielens}, {Peeters},
  \& {Jones}}]{Galliano2008}
{Galliano}, F., {Madden}, S.~C., {Tielens}, A.~G.~G.~M., {Peeters}, E., \&
  {Jones}, A.~P. 2008, \apj, 679, 310

\bibitem[{{Genzel} {et~al.}(1998){Genzel}, {Lutz}, {Sturm}, {Egami}, {Kunze},
  {Moorwood}, {Rigopoulou}, {Spoon}, {Sternberg}, {Tacconi-Garman}, {Tacconi},
  \& {Thatte}}]{Genzel1998}
{Genzel}, R., {Lutz}, D., {Sturm}, E., {et~al.} 1998, \apj, 498, 579

\bibitem[{{Haan} {et~al.}(2011){Haan}, {Armus}, {Laine}, {Charmandaris},
  {Smith}, {Schweizer}, {Brandl}, {Evans}, {Surace}, {Diaz-Santos}, {Beirao},
  {Murphy}, {Stierwalt}, {Hibbard}, {Yun}, \& {Jarrett}}]{Haan2011}
{Haan}, S., {Armus}, L., {Laine}, S., {et~al.} 2011, ArXiv e-prints

\bibitem[{{Harris} {et~al.}(2010){Harris}, {Rejkuba}, \& {Harris}}]{Harris2010}
{Harris}, G.~L.~H., {Rejkuba}, M., \& {Harris}, W.~E. 2010, \pasa, 27, 457

\bibitem[{{Haynes} {et~al.}(2010){Haynes}, {Cannon}, {Skillman}, {Jackson}, \&
  {Gehrz}}]{Haynes2010}
{Haynes}, K., {Cannon}, J.~M., {Skillman}, E.~D., {Jackson}, D.~C., \& {Gehrz},
  R. 2010, \apj, 724, 215

\bibitem[{{Israel}(1998)}]{Israel1998}
{Israel}, F.~P. 1998, \aapr, 8, 237

\bibitem[{{Kainulainen} {et~al.}(2009){Kainulainen}, {Alves}, {Beletsky},
  {Ascenso}, {Kainulainen}, {Amorim}, {Lima}, {Marques}, {Moitinho},
  {Pinh{\~a}o}, {Rebord{\~a}o}, \& {Santos}}]{Kainulainen2009}
{Kainulainen}, J.~T., {Alves}, J.~F., {Beletsky}, Y., {et~al.} 2009, \aap, 502,
  L5

\bibitem[{{Kennicutt} {et~al.}(2003){Kennicutt}, {Armus}, {Bendo}, {Calzetti},
  {Dale}, {Draine}, {Engelbracht}, {Gordon}, {Grauer}, {Helou}, {Hollenbach},
  {Jarrett}, {Kewley}, {Leitherer}, {Li}, {Malhotra}, {Regan}, {Rieke},
  {Rieke}, {Roussel}, {Smith}, {Thornley}, \& {Walter}}]{Kennicutt2003}
{Kennicutt}, Jr., R.~C., {Armus}, L., {Bendo}, G., {et~al.} 2003, \pasp, 115,
  928

\bibitem[{{Kraft} {et~al.}(2000){Kraft}, {Forman}, {Jones}, {Kenter}, {Murray},
  {Aldcroft}, {Elvis}, {Evans}, {Fabbiano}, {Isobe}, {Jerius}, {Karovska},
  {Kim}, {Prestwich}, {Primini}, {Schwartz}, {Schreier}, \&
  {Vikhlinin}}]{Kraft2000}
{Kraft}, R.~P., {Forman}, W., {Jones}, C., {et~al.} 2000, \apjl, 531, L9

\bibitem[{{Krajnovi{\'c}} {et~al.}(2007){Krajnovi{\'c}}, {Sharp}, \&
  {Thatte}}]{Krajnovic2007}
{Krajnovi{\'c}}, D., {Sharp}, R., \& {Thatte}, N. 2007, \mnras, 374, 385

\bibitem[{{Laurent} {et~al.}(2000){Laurent}, {Mirabel}, {Charmandaris},
  {Gallais}, {Madden}, {Sauvage}, {Vigroux}, \& {Cesarsky}}]{Laurent2000}
{Laurent}, O., {Mirabel}, I.~F., {Charmandaris}, V., {et~al.} 2000, \aap, 359,
  887

\bibitem[{{Leeuw} {et~al.}(2002){Leeuw}, {Hawarden}, {Matthews}, {Robson}, \&
  {Eckart}}]{Leeuw2002}
{Leeuw}, L.~L., {Hawarden}, T.~G., {Matthews}, H.~E., {Robson}, E.~I., \&
  {Eckart}, A. 2002, \apj, 565, 131

\bibitem[{{Ma} {et~al.}(1998){Ma}, {Arias}, {Eubanks}, {Fey}, {Gontier},
  {Jacobs}, {Sovers}, {Archinal}, \& {Charlot}}]{Ma1998}
{Ma}, C., {Arias}, E.~F., {Eubanks}, T.~M., {et~al.} 1998, \aj, 116, 516

\bibitem[{{Marconi} {et~al.}(2000){Marconi}, {Schreier}, {Koekemoer},
  {Capetti}, {Axon}, {Macchetto}, \& {Caon}}]{Marconi2000}
{Marconi}, A., {Schreier}, E.~J., {Koekemoer}, A., {et~al.} 2000, \apj, 528,
  276

\bibitem[{{Mirabel} {et~al.}(1999){Mirabel}, {Laurent}, {Sanders}, {Sauvage},
  {Tagger}, {Charmandaris}, {Vigroux}, {Gallais}, {Cesarsky}, \&
  {Block}}]{Mirabel1999}
{Mirabel}, I.~F., {Laurent}, O., {Sanders}, D.~B., {et~al.} 1999, \aap, 341,
  667

\bibitem[{{Netzer} {et~al.}(2007){Netzer}, {Lutz}, {Schweitzer}, {Contursi},
  {Sturm}, {Tacconi}, {Veilleux}, {Kim}, {Rupke}, {Baker}, {Dasyra},
  {Mazzarella}, \& {Lord}}]{Netzer2007}
{Netzer}, H., {Lutz}, D., {Schweitzer}, M., {et~al.} 2007, \apj, 666, 806

\bibitem[{{Peeters} {et~al.}(2004){Peeters}, {Allamandola}, {Hudgins}, {Hony},
  \& {Tielens}}]{Peeters2004}
{Peeters}, E., {Allamandola}, L.~J., {Hudgins}, D.~M., {Hony}, S., \&
  {Tielens}, A.~G.~G.~M. 2004, in Astronomical Society of the Pacific
  Conference Series, Vol. 309, Astrophysics of Dust, ed. {A.~N.~Witt,
  G.~C.~Clayton, \& B.~T.~Draine}, 141--+

\bibitem[{{Peeters} {et~al.}(2002){Peeters}, {Hony}, {Van Kerckhoven},
  {Tielens}, {Allamandola}, {Hudgins}, \& {Bauschlicher}}]{Peeters2002}
{Peeters}, E., {Hony}, S., {Van Kerckhoven}, C., {et~al.} 2002, \aap, 390, 1089

\bibitem[{{Quillen} {et~al.}(2006{\natexlab{a}}){Quillen}, {Bland-Hawthorn},
  {Brookes}, {Werner}, {Smith}, {Stern}, {Keene}, \&
  {Lawrence}}]{Quillen2006Shell}
{Quillen}, A.~C., {Bland-Hawthorn}, J., {Brookes}, M.~H., {et~al.}
  2006{\natexlab{a}}, \apjl, 641, L29

\bibitem[{{Quillen} {et~al.}(2008){Quillen}, {Bland-Hawthorn}, {Green},
  {Smith}, {Prasad}, {Alonso-Herrero}, {Cleary}, {Brookes}, \&
  {Lawrence}}]{Quillen2008}
{Quillen}, A.~C., {Bland-Hawthorn}, J., {Green}, J.~D., {et~al.} 2008, \mnras,
  384, 1469

\bibitem[{{Quillen} {et~al.}(2006{\natexlab{b}}){Quillen}, {Brookes}, {Keene},
  {Stern}, {Lawrence}, \& {Werner}}]{Quillen2006}
{Quillen}, A.~C., {Brookes}, M.~H., {Keene}, J., {et~al.} 2006{\natexlab{b}},
  \apj, 645, 1092

\bibitem[{{Rigopoulou} {et~al.}(1999){Rigopoulou}, {Spoon}, {Genzel}, {Lutz},
  {Moorwood}, \& {Tran}}]{Rigopoulou1999}
{Rigopoulou}, D., {Spoon}, H.~W.~W., {Genzel}, R., {et~al.} 1999, \aj, 118,
  2625

\bibitem[{{Schreier} {et~al.}(1998){Schreier}, {Marconi}, {Axon}, {Caon},
  {Macchetto}, {Capetti}, {Hough}, {Young}, \& {Packham}}]{Schreier1998}
{Schreier}, E.~J., {Marconi}, A., {Axon}, D.~J., {et~al.} 1998, \apjl, 499,
  L143+

\bibitem[{{Schutte} {et~al.}(1993){Schutte}, {Tielens}, \&
  {Allamandola}}]{Schutte1993}
{Schutte}, W.~A., {Tielens}, A.~G.~G.~M., \& {Allamandola}, L.~J. 1993, \apj,
  415, 397

\bibitem[{{Schweitzer} {et~al.}(2006){Schweitzer}, {Lutz}, {Sturm}, {Contursi},
  {Tacconi}, {Lehnert}, {Dasyra}, {Genzel}, {Veilleux}, {Rupke}, {Kim},
  {Baker}, {Netzer}, {Sternberg}, {Mazzarella}, \& {Lord}}]{Schweitzer2006}
{Schweitzer}, M., {Lutz}, D., {Sturm}, E., {et~al.} 2006, \apj, 649, 79

\bibitem[{{Smith} {et~al.}(2007){Smith}, {Draine}, {Dale}, {Moustakas},
  {Kennicutt}, {Helou}, {Armus}, {Roussel}, {Sheth}, {Bendo}, {Buckalew},
  {Calzetti}, {Engelbracht}, {Gordon}, {Hollenbach}, {Li}, {Malhotra},
  {Murphy}, \& {Walter}}]{Smith2007}
{Smith}, J.~D.~T., {Draine}, B.~T., {Dale}, D.~A., {et~al.} 2007, \apj, 656,
  770

\bibitem[{{Spoon}(2003)}]{Spoon2003}
{Spoon}, H.~W.~W. 2003, PhD thesis, Proefschrift, Rijksuniversiteit Groningen,
  2003

\bibitem[{{Spoon} {et~al.}(2007){Spoon}, {Marshall}, {Houck}, {Elitzur}, {Hao},
  {Armus}, {Brandl}, \& {Charmandaris}}]{Spoon2007}
{Spoon}, H.~W.~W., {Marshall}, J.~A., {Houck}, J.~R., {et~al.} 2007, \apjl,
  654, L49

\bibitem[{{Struve} {et~al.}(2010){Struve}, {Morganti}, {Oosterloo}, \&
  {Emonts}}]{Struve2010}
{Struve}, C., {Morganti}, R., {Oosterloo}, T.~A., \& {Emonts}, B.~H.~C. 2010,
  \pasa, 27, 390

\bibitem[{{Sturm} {et~al.}(2000){Sturm}, {Lutz}, {Tran}, {Feuchtgruber},
  {Genzel}, {Kunze}, {Moorwood}, \& {Thornley}}]{Sturm2000}
{Sturm}, E., {Lutz}, D., {Tran}, D., {et~al.} 2000, \aap, 358, 481

\bibitem[{{Tacconi-Garman} {et~al.}(2005){Tacconi-Garman}, {Sturm}, {Lehnert},
  {Lutz}, {Davies}, \& {Moorwood}}]{Tacconi-Garman2005}
{Tacconi-Garman}, L.~E., {Sturm}, E., {Lehnert}, M., {et~al.} 2005, \aap, 432,
  91

\bibitem[{{Teplitz} {et~al.}(2007){Teplitz}, {Desai}, {Armus}, {Chary},
  {Marshall}, {Colbert}, {Frayer}, {Pope}, {Blain}, {Spoon}, {Charmandaris}, \&
  {Scott}}]{Teplitz2007}
{Teplitz}, H.~I., {Desai}, V., {Armus}, L., {et~al.} 2007, \apj, 659, 941

\bibitem[{{Tran} {et~al.}(2001){Tran}, {Lutz}, {Genzel}, {Rigopoulou}, {Spoon},
  {Sturm}, {Gerin}, {Hines}, {Moorwood}, {Sanders}, {Scoville}, {Taniguchi}, \&
  {Ward}}]{Tran2001}
{Tran}, Q.~D., {Lutz}, D., {Genzel}, R., {et~al.} 2001, \apj, 552, 527

\bibitem[{{Valiante} {et~al.}(2007){Valiante}, {Lutz}, {Sturm}, {Genzel},
  {Tacconi}, {Lehnert}, \& {Baker}}]{Valiante2007}
{Valiante}, E., {Lutz}, D., {Sturm}, E., {et~al.} 2007, \apj, 660, 1060

\bibitem[{{van Diedenhoven} {et~al.}(2004){van Diedenhoven}, {Peeters}, {Van
  Kerckhoven}, {Hony}, {Hudgins}, {Allamandola}, \&
  {Tielens}}]{vanDiedenhoven2004}
{van Diedenhoven}, B., {Peeters}, E., {Van Kerckhoven}, C., {et~al.} 2004,
  \apj, 611, 928

\bibitem[{{Wiebe} {et~al.}(2011){Wiebe}, {Egorov}, \& {Lozinskaya}}]{Wiebe2011}
{Wiebe}, D.~S., {Egorov}, O.~V., \& {Lozinskaya}, T.~A. 2011, ArXiv e-prints

\bibitem[{{Yan} {et~al.}(2007){Yan}, {Sajina}, {Fadda}, {Choi}, {Armus},
  {Helou}, {Teplitz}, {Frayer}, \& {Surace}}]{Yan2007}
{Yan}, L., {Sajina}, A., {Fadda}, D., {et~al.} 2007, \apj, 658, 778

\end{thebibliography}
\end{document}